\documentclass[11pt]{article}

\usepackage{latexsym,amsmath,epsfig}

\DeclareFontFamily{OT1}{rsfs10}{}
\DeclareFontShape{OT1}{rsfs10}{m}{n}{ <-> rsfs10 }{}
\DeclareMathAlphabet{\mathscript}{OT1}{rsfs10}{m}{n}

\numberwithin{equation}{section}

\newcommand{\be}{\begin{equation}}
\newcommand{\ee}{\end{equation}}
\newcommand{\nn}{\nonumber}
\newcommand{\bea}{\begin{eqnarray}}
\newcommand{\eea}{\end{eqnarray}}

\newcommand{\ns}{\normalsize}
\newcommand{\pt}{\partial}

\def\a{\alpha}
\def\b{\beta}
\def\g{\gamma}

\def\d{\delta}
\def\e{\epsilon}

\def\f{\phi}

\def\k{\kappa}

\def\m{\mu}
\def\n{\nu}

\def\p{\pi}

\def\r{\rho}

\def\t{\tau}

\def\cM{{\cal M}}
\def\cN{{\cal N}}

\begin{document}


\begin{titlepage}

\vspace{-3cm}

\title{
   \hfill{\ns OUTP-99-09P, UPR-831T, PUPT-1837\\}
   \hfill{\ns hep-th/9902071} \\[3em]
   {\huge Boundary Inflation}\\[1em]}
\author{
{\ns\large Andr\'e Lukas$^1$, Burt A.~Ovrut$^2$
            and Daniel Waldram$^3$} \\[0.8em]
   {\it\ns $^1$Department of Physics, Theoretical Physics, 
     University of Oxford} \\[-0.2em]
      {\it\ns 1 Keble Road, Oxford OX1 3NP, United Kingdom} \\[0.2em]   
   {\it\ns $^2$Department of Physics, University of Pennsylvania} \\[-0.2em]
      {\it\ns Philadelphia, PA 19104--6396, USA}\\[0.5em]
   {\it\ns $^3$Department of Physics, Joseph Henry Laboratories,} \\[-0.2em]
      {\it\ns Princeton University, Princeton, NJ 08544, USA}}
\date{}

\maketitle

\begin{abstract}
Inflationary solutions are constructed in a specific five--dimensional
model with boundaries motivated by heterotic M--theory. We
concentrate on the case where the vacuum energy is provided by
potentials on those boundaries. It is pointed out that the presence of
such potentials necessarily excites bulk Kaluza--Klein modes. We
distinguish a linear and a non--linear regime for those modes. In the
linear regime, inflation can be discussed in an effective
four--dimensional theory in the conventional way. We lift a
four--dimensional inflating solution up to five dimensions where it
represents an inflating domain wall pair. This shows explicitly
the inhomogeneity in the fifth dimension. We also demonstrate the
existence of inflating solutions with unconventional properties in the
non--linear regime. Specifically, we find solutions with and without
an horizon between the two boundaries. These solutions have certain
problems associated with the stability of the additional dimension
and the persistence of initial excitations of the Kaluza--Klein modes. 
\end{abstract}

\thispagestyle{empty}

\end{titlepage}


\section{Introduction}

Two important theoretical developments, the advent of M--theory and
the discovery of branes have recently stimulated new ideas in early
universe cosmology. There has been considerable activity on various
cosmological aspects of M--theory over the past
two years~\cite{cos1}--\cite{lk}. The cosmology of Ho\v rava--Witten
theory~\cite{hw1,hw2,hor,w}, which describes M--theory on the orbifold
$S^1/Z_2$, however, is much less studied so
far~\cite{b1,b2,cosm,cosm1,real}. This theory describes the
strong coupling dual of the $E_8\times E_8$ heterotic string and is,
therefore, of great importance for M--theory particle phenomenology.
Clearly, this property makes it a very interesting starting point for
cosmology as well. 

Various aspects of branes might be important in early universe
cosmology, such as their ability to smooth out
singularities~\cite{LW,cos2,bfs,mr,bfm}
and their thermodynamical properties~\cite{rama,cp,bkr,abkr}. Most obviously,
however, they play a role in the cosmology of particle physics models
that have branes in their vacuum structure and, more specifically,
that lead to low--energy theories arising from the worldvolumes of branes. 
Such models appear in the context of brane boxes~\cite{hz,hu}, type I
string theory~\cite{ibanez,lykken,kt,biq} and M--theory on
$S^1/Z_2$~\cite{losw,losw1,dlow,dlow1}. A
characteristic feature of many of those models is the possibility of
one or more compact dimensions being large compared to the fundamental
length scale of the theory. Such a situation can be described by a
Kaluza--Klein theory with gravity and possibly other fields in the
bulk coupled to a four--dimensional ``brane--like'' object with the
standard model fields on its worldvolume. A wide spectrum of scales has been
proposed for such models. These range from fundamental scales around the
GUT scale with the energy scale associated with the additional
dimensions being an order of magnitude or so smaller, to
models with a fundamental scale of order a TeV with macroscopic
additional dimensions~\cite{dim1}. It is clearly interesting to explore
the cosmology of these models and, recently, some work in this
direction~\cite{add,mr,cosm,real,ddgr} has been done.

In this paper, we would like to study the important issue of how
inflation relates to these new theoretical ideas. For recent
related work on inflation see~\cite{ekoy,ly1,kl,dt,lk,bsch}. Rather than
presenting a general scenario, we will concentrate on a specific model which
incorporates the M--theory as well as the brane aspects.
This model can be interpreted as a part of the five--dimensional
effective action of M--theory on
$S^1/Z_2$~\cite{losw,elpp,losw1,elpp1}
obtained by reducing the 11--dimensional theory on a Calabi--Yau three--fold.
The five--dimensional space of this theory has the structure
$\cM_5=S^1/Z_2\times \cM_4$ and contains two four--dimensional orbifold
fixed planes (or boundaries) $\cM_4^{(1)}$ and $\cM_4^{(2)}$. It consists
of gauged $\cN =1$ supergravity plus vector-- and hypermultiplets in
the bulk coupled to $\cN =1$ theories with gauge and chiral multiplets
on the orbifold fixed planes. The vacuum solution of this
theory~\cite{losw} is a BPS double three--brane (domain wall) with the
three--brane worldvolumes identified with the orbifold planes. Upon
reduction to four dimensions on this vacuum solution, one arrives at
an $\cN =1$ supergravity theory which is a candidate for a realistic particle
physics model from M--theory. The hidden and observable fields in this
model arise from the ``three--brane orbifold planes''.
The theory, therefore, allows us to study cosmology in a potentially
realistic particle physics environment and provides a concrete realization
of the general idea of ``getting the standard model from a brane''.
Simple cosmological solutions of this theory have been found in 
ref.~\cite{cosm,real}. In ref.~\cite{dlow,dlow1} non--perturbative
vacua of heterotic M--theory containing five--branes have been
constructed. In the five--dimensional effective theory, these
five--branes appear as three--branes which, in addition to the two
orbifold planes, are coupled to the bulk. It would clearly be
interesting to study cosmological solutions of these more general
theories with additional three--branes. In this paper, however, we
restrict ourselves to the simple setting of two orbifold planes.

For the application to cosmology, we have (consistently) truncated this
theory to a minimal field content suitable for inflationary models. 
Specifically, in the bulk we have kept gravity and a scalar field $\f$
(the volume modulus of the internal Calabi--Yau space) with
a potential $V$ of non--perturbative origin. In addition, on each
orbifold plane we have also kept a scalar field $\f_i$ with a potential
$V_{4i}$. The theory is characterized by three scales, namely the
fundamental scale $M_5$ of the five--dimensional theory, the separation $R$
of the orbifold planes and a scale $v$ that sets the height of certain
explicit potentials for the bulk field $\f$. These potentials are
responsible for the existence of the domain wall solution. 

Hence, we have a very simple setting with one additional dimension and
one ``candidate inflaton'' with potential in each part of the
space. In addition to the M--theory relation, we, clearly, also have a
simple starting point to study inflation in the general context of 
models with large additional dimensions.

The goal of this paper is not to construct explicit
inflationary models by choosing specific potentials for the scalar
fields in the theory. Rather, we are interested in how the specific
structure of the theory, that is, the coupling of the
five--dimensional bulk to four--dimensional boundary theories, effects
inflation. We distinguish two different types of inflation, which we
call bulk (or moduli) and boundary (or matter field) inflation. For
bulk inflation the vacuum energy is predominantly provided by the bulk
potential $V$, whereas for boundary inflation the boundary potentials
$V_{4i}$ dominate. In this paper, we concentrate on the second case of
boundary inflation. This option is particularly interesting in that it
directly relates to the presence of the characteristic boundary
theories. Moreover, inflation from matter fields seems to be in better
accord with current directions in four--dimensional inflationary model
building~\cite{lr} than modular inflation.

\vspace{0.4cm}

Let us summarize our main results. One of the main themes of this
paper is that energy density on the orbifold planes provides
source terms localized on the fixed points in the additional dimension
and, hence, excites
bulk fields. In particular, this applies to vacuum energy on these
planes as needed for boundary inflation. Our first conclusion is
that boundary inflation is necessarily inhomogeneous in the
additional dimension, or, in other words, excites Kaluza--Klein
modes. The magnitude of those excitations is controlled by the
dimensionless parameter $\e_i=V_{4i}R/M_5^3$. For $|\e_i|\ll 1$ the
excitations can be described by linearized gravity. This approximation
breaks down if $|\e_i|\gg 1$. One then has to use the full non--linear
theory. Using the COBE normalization and the typical magnitude of
heterotic M--theory scales, we argue that inflation in this theory
may take place in both regimes. Interestingly, if the orbifold and
Calabi--Yau scales during inflation are at their physical
values~\cite{w}, the theory becomes linear for $V_{4i}^{1/4}\ll 6\cdot
10^{16}$ GeV, right below the COBE scale.

The proportionality of $\e_i$ to the
size $R$ of the additional dimension can be understood from the linear
behaviour of the one--dimensional Green's function. For more than one
additional dimension, the Green's function is logarithmic or
follows an inverse power law. As a consequence, in those cases, the
theory is always in the linear regime. The case of heterotic M--theory
with one large dimension is, therefore, the only one where inflation
may take place in the non--linear regime.

In the linear regime, we can compute a sensible four--dimensional
effective theory by integrating out the Kaluza--Klein modes induced by
the boundary sources. We show explicitly how this leads to corrections
in the four--dimensional theory. Our basic statement is that, in the
linear regime, inflation can essentially be treated in the effective
four--dimensional $\cN =1$ supergravity theory. Nevertheless, to get
a physical picture, we find it useful to lift a generic
four--dimensional inflating solution up to a five--dimensional one.
This five--dimensional solution represents a pair of inflating domain
wall three--branes and it has inhomogeneities in the additional
dimension caused by the boundary potentials. On the other hand,
initial inhomogeneities not induced by boundary sources are damped
away in the linear regime due to the inflationary expansion and
should not play any role.

The situation is quite different in the non--linear regime, $\e_i\gg
1$, where one has to solve the full five--dimensional theory. To do
so, we assume that the bulk scalar field $\f$ has been stabilized by
its potential and the boundary potentials allow for slow roll
behaviour of the boundary scalars. Under these assumptions we find,
in a first attempt, a simple solution by separation of variables
that exhibits inflation. This solution represents the heterotic
M--theory version of an old four--dimensional domain wall
solution~\cite{vil,is}, recently advocated~\cite{kl} in a somewhat
different approach to brane inflation. Both boundaries expand in a de
Sitter--like manner with the Hubble parameters $H_i$ related to the
potentials in an unconventional way, $H_i\sim |V_{4i}|M_5^{-3}$. The
physical size $R_{\rm phys}$ of the additional dimension is constant
in time and fixed in terms of the boundary potentials. For $H_1$ and
$H_2$ of the same order, $H_1\sim H_2\sim H$, one has
$R_{\rm phys}\sim 1/H$. The Hubble parameter, therefore, equals the
orbifold energy scale during inflation. We find solutions with and without an
horizon at some point on the orbifold. The solutions without horizon
require potentials with opposite sign satisfying $V_{41}+V_{42}<0$.
Signals travel from one boundary to the other in a finite time and
every signal emitted somewhere in the bulk will eventually reach one of the
boundaries. For the solutions with horizon one needs both
potentials to be positive, $V_{4i}>0$. In this case, the two
boundaries are causally decoupled. A signal emitted on one boundary
will never reach the other one.

In a second approach, we then find the general solution of the model
without assuming separation of variables. This is done exploiting
the similarity of our equations to those of two--dimensional dilaton
gravity~\cite{callan}. We recover
the previous inflating solution as a special case if a certain
continuous parameter in the general solution is set to zero. For all
other values of this parameter, however, the solution is
non--inflating and has a collapsing orbifold. This indicates an
instability of the solution which might be cured by stabilizing the
modulus of the additional dimension. The construction of a
viable inflationary background in the non--linear regime is, therefore,
tied to the question on how precisely such a stabilization is realized.
We discuss various options and their consequences in
our context. Another problem with the inflating solution which is
made visible by its generalization is the appearance of an arbitrary
periodic function in the solution. This function encodes the initial
inhomogeneities in the additional dimension. Unlike in the linear
case, here these inhomogeneities are not damped away. This seems to
be in contradiction with the inflationary paradigm that all initial
information should be wiped out. On the other hand, if sufficiently
small, these inhomogeneities may lead to interesting predictions.
Given those problems, we point out that conventional inflation in the
linear regime remains a perfectly viable option for heterotic
M--theory. For models with more than one large dimension it is
the only possibility. 

\vspace{0.4cm}

Based on the results of this paper, we would like to propose three
scenarios for inflation in heterotic M--theory.
\begin{itemize}
\item The orbifold and the Calabi--Yau scales during inflation are at
the specific values that at low-energy lead to coupling
unification. In this case, the
theory becomes linear for boundary potentials satisfying $V_{4i}^{1/4}\ll
V_{\rm lin}^{1/4}\simeq 6\cdot 10^{16}$ GeV. At the unification point
the Calabi--Yau scale and the fundamental 11--dimensional Planck
scale are also of the order $10^{16}$ GeV. This theory undergoes a
transition from a pure M--theory regime at energies above $10^{16}$
GeV (where no description in terms of 11--dimensional supergravity
applies) directly to the linear regime. Inflation can then take place in the
conventional way, presented in this paper. In this scenario, the
energy density at the beginning of inflation is directly linked
to the fundamental scales of the theory and is, in this sense,
explained. It fits the COBE normalization $V_4^{1/4}\simeq
\varepsilon^{1/4}\, 6.7\cdot 10^{16}$ GeV if the slow--roll parameter
$\varepsilon$ is not too small, or, in other words, if the inflaton
potential is not too flat.
\end{itemize}
An alternative possibility is that the orbifold and Calabi--Yau scales
are not be at the coupling unification values during inflation.
Rather, they are such that the linear regime starts significantly
below $10^{16}$ GeV, $V_{\rm lin}^{1/4}\ll 10^{16}$ GeV.
In this case, there are two options.
\begin{itemize}
\item Inflation takes place in the non--linear regime and is based on
the corresponding solutions given in this paper. This option is
somewhat speculative as it depends on the successful stabilization of
the orbifold modulus. It would have very unconventional properties.
These include a linear relation between the Hubble
parameter and the potential and inhomogeneities in the orbifold
direction. Since space--time in this context is genuinely
five--dimensional, analyzing density fluctuations requires some care
and the standard equations may not apply. 
\item Non--linear inflation does not take place. This might happen for
a number of reasons. For example, the potentials might not have the
required properties, the initial conditions may not be appropriate or,
simply, non--linear inflation might not work at all. Inflation could
then start when the energy density drops below $V_{\rm lin}$ and the
linear regime is reached. This could be consistent with the COBE
normalization for a very small slow--roll parameter $\varepsilon$,
that is, a very flat inflaton potential.
\end{itemize}

\section{The action}

In this section, we would like to present the five--dimensional action
that we are going to use in this paper along with its most important
properties. This includes a discussion of its origin and
interpretation, its ``vacuum'' solution and the related
four--dimensional effective low--energy theory that is obtained as a
reduction on this vacuum solution. Making contact with the
four--dimensional theory is particularly useful, in our context,
whenever the relation to ``conventional'' four--dimensional inflation
is analyzed.

Our starting point is the five--dimensional action~\footnote{We have
changed somewhat the notation with respect to ref.~\cite{losw} to be
in better accord with conventions in cosmology. The scalar field $\f$
is related to the field $V$ of ref.~\cite{losw} by $V=e^\f$. The
constant $v$ was called $\a_0$ there.}
\bea
 S_5 &=& -\frac{1}{2\k_5^2}\left\{\int_{\cM_5}\sqrt{-g}\left[
         R+\frac{1}{2}\pt_\a\f\pt^\a\f +U(\f )\right]\right. \nn\\
      &&\qquad\qquad\left. +\sum_{i=1}^2\int_{\cM_4^{(i)}}
        \sqrt{-g}\left[\frac{1}{2}
        \pt_\m\f_i\pt^\m\f_i+U_i(\f_i ,\f )\right]\right\} \label{S5}
\eea
where the potentials are given by
\bea
 U(\f ) &=& \frac{1}{3}v^2e^{-2\f}+V(\f ) \label{U}\\
 U_i(\f_i,\f ) &=& \mp 2\sqrt{2}ve^{-\f}+V_i(\f_i,\f )\; .\label{Ui}
\eea
Here $\k_5$ is the five--dimensional Newton constant. Coordinates
$x^\a$ with indices $\a ,\b ,\g ,\dots = 0,\dots ,3,5$ are used for
the five--dimensional space $\cM_5$. We consider a space--time with structure
$\cM_5=S^1/Z_2\times \cM_4$ where $S^1/Z_2$ is an orbifold and $\cM_4$ a smooth
four--manifold. The coordinates $x^\m$ on $\cM_4$ are labelled by
indices $\m ,\n ,\r ,\dots = 0,\dots ,3$ while the remaining
coordinate $y\equiv x^5$ parameterizes the orbifold. It is chosen in
the range $y\in [-R ,R ]$ with the endpoints identified, where
$R=\p\r$ and $\r $ is the radius of the orbicircle. The
$Z_2$ symmetry acts as $y\rightarrow -y$ and
leaves two four--dimensional planes, characterized by $y_1=0$ and
$y_2=R$, fixed. These planes, separated by a distance $R$, are denoted by
$\cM_4^{(i)}$, where $i=1,2$. The action~\eqref{S5} describes
five--dimensional gravity plus a scalar field $\f$ with potential $U$
in the bulk coupled to four--dimensional theories on the orbifold
fixed planes each carrying a scalar field $\f_i$ with potential $U_i$.
The bulk fields have to be truncated in accordance with the $Z_2$
symmetry. Specifically, one should require
\bea
 \f (-y) &=& \f (y) \nn \\
 g_{\m\n}(-y) &=& g_{\m\n}(y) \nn \\
 g_{\m 5}(-y) &=& -g_{\m 5}(y) \\
 g_{55}(-y) &=& g_{55}(y) \nn \; .
\eea
Hence, $\f$, $g_{\m\n}$, $g_{55}$ are even under the $Z_2$ symmetry
while $g_{\m 5}$ is odd. Also note that the $y$--derivative of an even
field is odd and vice versa. Whereas even fields are continuous across
the orbifold planes, odd field jump from a certain value on one side
to its negative on the other side. An alternative and equivalent
way to think about the
five--dimensional space, in contrast to the orbifold picture which we
have used so far, is the boundary picture. In this picture, the
coordinate $y$ is restricted to one half of the circle, $y\in I=[0,R]$
and five--dimensional space--time is written as $\cM_5 =
I\times\cM_4$. The orbifold fixed planes then turn into the boundaries
of this five--dimensional space. We will sometimes find it more
convenient to work in this boundary picture.

The potentials~\eqref{U}
and \eqref{Ui} have been split into explicit, exponential potentials
for $\f$ with a height set by the constant $v$ and potentials $V$ and $V_i$
which, at this point, are arbitrary. The $-$ sign in \eqref{Ui} refers
to the plane $i=1$, the $+$ sign to the plane $i=2$. The reason for
writing the potentials in this form will be explained shortly. It is
useful to collect the mass dimensions of the various objects that we
have introduced. In five dimensions, the Newton constant $\k_5^{-2}$ has
dimension three and we write
\begin{equation}
 \k_5^{-2}=M_5^3\label{M5}\; .
\end{equation}
Here $M_5$ is the ``fundamental'' scale of the five--dimensional theory.
In order to make subsequent equations simpler, we have pulled the
Newton constant in front of the complete action. Hence, the bulk potentials
$U$, $V$ have dimension two whereas the boundary potentials $U_i$,
$V_i$ and the constant $v$ have dimension one. 

\vspace{0.4cm}

Let us now discuss the interpretation of the
action~\eqref{S5}--\eqref{Ui} in terms of M--theory. More
specifically, we will consider Ho\v rava--Witten theory; that is,
11--dimensional supergravity on the space $S^1/Z_2\times \cM_{10}$
where $\cM_{10}$ is a smooth ten--dimensional manifold. The two
10--dimensional orbifold fixed planes of this theory carry additional
degrees of freedom that couple to the bulk supergravity, namely two
$E_8$ gauge multiplets, one
on each plane. Now consider reducing this theory on
a Calabi--Yau three--fold assuming that the radius of the Calabi--Yau
space is smaller than the orbifold radius. For the present values of
these radii, such a relation is suggested by coupling constant
unification~\cite{w}. We then arrive at a sensible five--dimensional
theory on the space--time $\cM_5=S^1/Z_2\times \cM_4$, where the two
four--dimensional fixed planes of the orbifold result from the
original 10--dimensional planes. For the standard embedding of the spin
connection into one of the $E_8$ gauge groups, this effective action has been
computed in ref.~\cite{losw,elpp,losw1}. The generalization to
non--standard embedding has been described in ref.~\cite{nse}.
It turns out that the bulk theory is a five--dimensional $\cN =1$ gauged
supergravity coupled to vector-- and hypermultiplets. This bulk theory
is coupled to two four--dimensional $\cN =1$ theories that reside on
the now four--dimensional orbifold planes. More specifically, these
boundary theories contain gauge multiplets as well as chiral (gauge
matter) multiplets. Upon appropriate reduction on the orbifold to four
dimensions (in a way to be specified below), one obtains a candidate
for a ``realistic'' $\cN =1$ supergravity theory with the observable sector
coming from one plane and the hidden sector from the other.

The action~\eqref{S5} is a ``universal'' version of this five--dimensional
effective theory truncated to the field content that seems essential
to discuss cosmology. Let us try to make this statement more precise
by explaining the meaning of the various objects in
\eqref{S5}--\eqref{Ui} in terms of the underlying 11--dimensional
theory. In terms of the 11--dimensional Newton constant $\k$ and the
Calabi--Yau volume $v_{\rm CY}$ the five--dimensional Newton constant
is given by
\begin{equation}
 \k_5^2 =\frac{\k^2}{v_{\rm CY}}\; .\label{k5}
\end{equation}
The bulk field $\f$ is simply the modulus associated with the
Calabi--Yau volume such that the ``physical'' volume is given by
$e^\f v_{\rm CY}$. Clearly, the Calabi--Yau space has, in general,
many more moduli associated with its shape and complex structure. For
simplicity, we have kept the volume modulus only, since it is the
geometrical modulus common to all Calabi--Yau spaces. It is in this
sense that we are referring to the action as universal. Also, for
simplicity, we have
dropped a number of other bulk fields such as the axions associated with
$\f$ and the vector fields in the vector multiplets. All these fields
can be consistently set to zero in the full equations of motion.
Therefore, our solutions will be solutions of the complete action
of heterotic M--theory as well. We have, however, kept
a feature of the action that arises from the gauging of the bulk
supergravity and is essential to discuss cosmology; namely, the
explicit bulk potential for $\f$ in eq.~\eqref{U}. The constant $v$ in
this potential is given by
\begin{equation}
 v = -\frac{\p}{\sqrt{2}}\left(\frac{\k}{4\p}\right)^{2/3}
     \frac{n}{v_{\rm CY}^{2/3}}\label{v}
\end{equation}
where $n$ is an instanton number related to the tangent space and
gauge instantons in the internal Calabi--Yau space. We have added
another potential $V(\f )$ in the bulk which one expects to
arise from non--perturbative effects like internally wrapped
membranes. The explicit computation of this potential for M--theory is
not very well understood at present (see however \cite{w1,wdg}). It
is, however, clear that such a potential is eventually needed to
stabilize the moduli. Given the lack of theoretical knowledge, we will
not assume any specific origin for $V$ but, rather, try to specify what
properties it should be required to have from a cosmological viewpoint.

Let us now move on to the four--dimensional theories on the orbifold
planes and their M--theory origin. As already mentioned, these planes
arise directly from the 10--dimensional planes of Ho\v rava--Witten
theory upon reduction on the Calabi--Yau space. On each of these
planes, we have introduced a scalar field $\f_i$ with potential $U_i$
that represents the scalar partners of the matter fields. In fact, for
non--standard embedding such matter fields generally arise on both
planes. Of course, restricting to one scalar field on each plane is a
tremendous simplification from the particle physics point of
view. Cosmologically, however, it seems reasonable to choose such a
model with one ``candidate inflaton'' on each orbifold plane,
especially in a first study of inflation in such models. Although the
potentials $V_i$ in eq.~\eqref{Ui} are known in principle for a given
Calabi--Yau compactification, here we will not attempt to be more specific
about their form. Instead of going into such detailed questions of
model building, we will assume that they have the cosmologically
desired properties. There is another, explicit part of the boundary
potential in eq.~\eqref{Ui} which depends on the projection of the
bulk field $\f$ onto the orbifold planes. These potentials
originate directly from Ho\v rava--Witten theory and are related to the
explicit bulk potential for $\f$. Note that, in particular, the height
is set by the same constant $v$, defined in eq.~\eqref{v}. As we will see,
they support a three--brane domain wall solution of the
five--dimensional theory. In the following, we, therefore, refer to
them as domain wall potentials. Of course,
there will be gauge multiplets on the orbifold planes as well which we
have not written in~\eqref{S5}. They could play a role in cosmology
via gaugino condensation. However, we will not consider
this explicitly in the present paper. To summarize, the
action~\eqref{S5}--\eqref{Ui} can be viewed as part of the
five--dimensional effective theory of heterotic M--theory and it
contains the basic cosmologically relevant ingredients of this theory. 

\vspace{0.4cm}

The M--theory context is not obligatory
here. Instead, the action~\eqref{S5} could be viewed in the general
context of theories with large additional dimensions where the
standard model arises from a brane worldvolume. In fact, in this
context, our action is about the simplest appropriate to study
inflationary cosmology. We have one ``large'' dimension (the orbifold)
and two four--dimensional brane--like theories corresponding to an
observable and a hidden sector. In addition to bulk gravity, we have a
bulk scalar field with potential and a scalar field with potential on
each plane; in other words, a minimal setting for inflation. The
explicit potentials in \eqref{U}, \eqref{Ui} that originate from
M--theory can always be switched off by setting $v=0$, if desired. Our
results are, therefore, not limited by their presence. What does limit
our results is the presence of only one additional dimension,
something that is appropriate for heterotic M--theory but not
necessarily otherwise. We will comment on the modification of our results
in the case of two or more large dimensions as we proceed.

\vspace{0.4cm}

For later reference, let us collect the equations of motion derived
from the action~\eqref{S5}. We have the Einstein equation
\begin{equation}
 R_{\a\b}-\frac{1}{2}g_{\a\b}R = T_{\a\b}+\sum_{i=1}^2T_{\a\b}^{(i)}
                                 \d (y-y_i)
\end{equation}
with the bulk and boundary energy momentum tensors
\bea
 T_{\a\b} &=& -\frac{1}{2}\left(\pt_\a\f\pt_\b\f
              -\frac{1}{2}g_{\a\b}(\pt\f)^2\right)+\frac{1}{2}g_{\a\b}U\\
 T_{\m\n}^{(i)} &=&-\frac{1}{2}g_{55}^{-1/2}\left(\pt_\m\f_i\pt_\n\f_i
                   -\frac{1}{2}g_{\m\n}(\pt\f_i)^2-g_{\m\n}U_i\right)\\
 T_{\m 5}^{(i)} &=& 0\; ,\qquad T_{55}^{(i)} = 0\; .
\eea
For the scalar fields we find
\bea
 \Box_5\f-\pt_\f U-g_{55}^{-1/2}\sum_{i=1}^2\pt_\f U_i\d (y-y_i) &=& 0\\
 \Box_4\f_i-\pt_{\f_i}V_i &=& 0
\eea
where $\Box_5$ and $\Box_4$ are the Laplacians associated with the
five--dimensional metric $g_{\a\b}$ and its four--dimensional part
$g_{\m\n}$ projected onto one of the orbifold planes. 

Which types of solutions to these equations are we interested in? For
a cosmological solution, one would like to have a three--dimensional
maximally symmetric subspace which, for simplicity, we take to be
flat. Studying open and closed universes would be clearly interesting
as well. Later, we will distinguish between a linear and a non--linear
case. While the generalization to include spatial curvature is
straightforward in the linear case, the non--linear case is
significantly more complicated. Still, we do not expect our main
conclusions to depend on the choice of spatial curvature. Clearly,
the maximally symmetric subspace cannot contain the orbifold. Hence, the
solutions are independent of ${\bf x}=(x^1,x^2,x^3)$. They can,
however, depend on the time $\t =x^0$ as well as on the orbifold
coordinate $y$. As we will see in a moment, the presence of fields and
potentials on the boundaries in fact forces us to consider orbifold
dependence. Therefore, we start with the ansatz
\bea
 ds_5^2 &=& -e^{2\n}d\t^2+e^{2\a}d{\bf x}^2+e^{2\b}dy^2 \\
 \n &=& \n (\t ,y) \\
 \a &=& \a (\t ,y) \\
 \b &=& \b (\t ,y) \\
 \f &=& \f (\t ,y) \\
 \f_i &=& \f_i(\t )\; .
\eea
Here $\a$ is the scale factor of the three--dimensional universe and
$\b$ is an additional scale factor that measures the orbifold
size. In some cases, we will choose the conformal gauge $\n =\b$ in
the above metric. Note that we can choose this gauge in the $(\t ,y)$
subspace in such a way that the boundaries are mapped to hypersurfaces
$y=$ const. Usually, then, we can shift the boundaries back to $y=0,R$
so that our conventions for the coordinate system remain intact. A
special case occurs if a boundary has been ``mapped to infinity'' due
to a singularity in the reparameterization. To be able to deal with
this case, we keep $\n$ arbitrary in the subsequent formulae.

The equations of motion for the above ansatz are given by
\begin{multline}
 3e^{-2\n}(\dot{\a}^2+\dot{\a}\dot{\b})-3e^{-2\b}(\a ''-\a '\b '+2{\a '}^2)=\\
 \frac{1}{4}e^{-2\n}\dot{\f}^2+\frac{1}{4}e^{-2\b}{\f '}^2+\frac{1}{2}U
 +\frac{1}{2}e^{-\b}\sum_{i=1}^2\left[\frac{1}{2}e^{-2\n}\dot{\f}_i^2
 +U_i\right]\d (y-y_i) \label{eomf}
\end{multline}
\begin{multline}
e^{-2\n}(2\ddot{\a}+\ddot{\b}+(3\dot{\a}+2\dot{\b}-2\dot{\n})\dot{\a}
+(\dot{\b}-\dot{\n})\dot{\b})-\\
e^{-2\b}(2\a ''+\n ''+(3\a '-2\b '+2\n ')\a '+(\n '-\b ')\n ')\\
=-\frac{1}{4}e^{-2\n}\dot{\f}^2
 +\frac{1}{4}e^{-2\b}{\f '}^2+\frac{1}{2}U
-\frac{1}{2}e^{-\b}\sum_{i=1}^2\left[\frac{1}{2}e^{-2\n}\dot{\f}_i^2
 -U_i\right]\d (y-y_i) 
\end{multline}
\bea
3e^{-2\n}(\ddot{\a}-\dot{\n}\dot{\a}+2\dot{\a}^2)-3e^{-2\b}({\a
'}^2+\n '\a ') &=&
-\frac{1}{4}e^{-2\n}\dot{\f}^2-\frac{1}{4}e^{-2\b}{\f
'}^2+\frac{1}{2}U \label{eom55}\\
3(\dot{\a}'+\dot{\a}\a '-\dot{\a}\n '-\dot{\b}\a
')&=&-\frac{1}{2}\dot{\f}\f ' \label{eomr}\\
e^{-2\n}(\ddot{\f}+(3\dot{\a}+\dot{\b}-\dot{\n})\dot{\f})-
e^{-2\b}(\f ''+(3\a '-\b '+\n ')\f ') &=&
-\pt_\f U-e^{-\b}\sum_{i=1}^2\pt_\f U_i\d (y-y_i)\\
\ddot{\f}_i+(3\dot{\a}-\dot{\n})\dot{\f}_i+e^{2\n}\pt_{\f_i}V_i &=& 0
 \label{eomm}
\eea
where the dot and the prime denote the derivatives with respect to
$\t$ and $y$. Alternatively, one can formulate these equations in the
boundary picture. Then all $\d$ function terms in the above equations
have to be dropped and are replaced by the following boundary conditions
\bea
 e^{-\b}\a '\left.\right|_{y=y_i} &=& \mp\frac{1}{12}\left[\frac{1}{2}e^{-2\n}
   \dot{\f}_i^2+ U_i\right]_{y=y_i} \label{alphab}\\
 e^{-\b}\n '\left.\right|_{y=y_i} &=& \mp\frac{1}{12}\left[-\frac{5}{2}e^{-2\n}
   \dot{\f}_i^2+U_i\right]_{y=y_i} \label{betab}\\
 e^{-\b}\f '\left.\right|_{y=y_i} &=&\pm\frac{1}{2}\left[\pt_\f U_i
    \right]_{y=y_i}\; .\label{eoml}
\eea
Here the upper (lower) sign applies to the first boundary at $y=y_1$
(the second boundary at $y=y_2$). The $\d$--function terms, or
equivalently the right hand sides of the boundary conditions, are
non--vanishing if there is any kinetic or potential energy on the
boundaries. Hence, in this most interesting case, the solution is
necessarily inhomogeneous in the orbifold coordinate $y$.

\vspace{0.4cm}

For the application to cosmology, one is also interested in the
four--dimensional effective action of~\eqref{S5} that is, roughly
speaking, valid when all energy scales are smaller than the
orbifold scale $1/R$. Normally, this action could be derived by a
simple truncation where all the bulk fields are taken to be
independent of the orbifold coordinate. While, in our case, this gives
the correct answer to lowest order, it neglects higher order
corrections that can be relevant. These corrections appear because,
strictly, we are not allowed to take the bulk fields independent of
the orbifold coordinate, thereby neglecting all contributions from
Kaluza--Klein modes. Instead, as we have just seen, every
non--vanishing term in the boundary actions produces an orbifold
dependence that needs to be integrated out and, typically, leads to
corrections to the effective action. Let us explain how this works in
our case, taking into account corrections up to the first non--trivial
(linear) order. For a more detailed account see~\cite{elten}. First,
we split the bulk fields as
\bea
 g_{\a\b} &=& \bar{g}_{\a\b}+\tilde{g}_{\a\b} \label{gsplit}\\
 \f &=& \bar{\f}+\tilde{\f} \label{fsplit}
\eea
into their orbifold average plus an orbifold--dependent
variation. Specifically, we have defined
\begin{equation}
 \bar{g}_{\a\b} = <g_{\a\b}>_5\; ,\qquad \bar{\f} = <\f >_5
\end{equation}
where $<.>_5$ denotes the average in the orbifold direction. Hence, the
average of the variations vanishes; that is,
\begin{equation}
 <\tilde{g}_{\a\b}>_5 = 0\; ,\qquad <\tilde{\f}>_5 = 0\; .
\end{equation}
The averaged fields are in one to one correspondence with the low energy
moduli fields. Aside from the modulus $\bar{\f}$, we have the orbifold
modulus $\bar{\b}$ and the four--dimensional Einstein--frame metric
$g_{4\m\n}$. The latter two quantities are related to the averaged
metric by
\bea
 e^{2\bar{\b}} &=& \bar{g}_{55} \\
 g_{4\m\n} &=& e^{\bar{\b}}\bar{g}_{\m\n}\; .
\eea
Note that we have no graviphoton zero mode since $g_{\m 5}$ is odd
under $Z_2$ and, hence, $\bar{g}_{\m 5}=0$. After expanding to linear
order in the variations, the equations of motion can be decomposed
into an averaged part and a part for the variations. The latter equations
have the following solutions in the boundary
picture~\footnote{This solution applies to the part of the boundary
stress energy which is homogeneous and constant in time on scales of
the order $R$. Slowly varying potential energy, the most important
case for this paper, typically satisfies this requirement.
For boundary processes on scales smaller than $R$ the associated bulk
fields are suppressed and decay exponentially away from the
boundary~\cite{low}.}
\bea
 \tilde{g}_{\a\b} &=& e^{2\bar{\b}}R\left[ P_1(z)\bar{S}_{\a\b}^{(1)}
                      +P_2(z)\bar{S}_{\a\b}^{(2)}\right] \label{gt}\\
 \tilde{\f} &=& \frac{1}{2}e^{\bar{\b}}R\left[P_1(z)\overline{\pt_\f U_1}
                      +P_2(z)\overline{\pt_\f U_2}\right] \label{ft}
\eea
where
\begin{equation}
 S_{\a\b}^{(i)}=T_{\a\b}^{(i)}-\frac{1}{3}g_{\a\b}g^{\g\d}T_{\g\d}^{(i)}
\end{equation}
is the modified boundary stress energy. The bar indicates that the
corresponding expressions are understood with the bulk fields replaced
by their zero modes. Furthermore, we have defined the
polynomials~\footnote{These polynomials are characterized by the
properties $<P_1(z)>_5=<P_2(z)>_5=0$, $P_1'(0)=1$, $P_1'(1)=0$,
$P_2'(0)=0$ and $P_2'(1)=-1$.}
\begin{equation}
 P_1(z)=-\frac{1}{2}z^2+z-\frac{1}{3}\; ,\qquad
 P_2(z) = -\frac{1}{2}z^2+\frac{1}{6}\label{pol}
\end{equation}
which depend on the normalized orbifold coordinate
\begin{equation}
 z=\frac{y}{R}\in [0,1]\; .
\end{equation}
This solution shows that each energy source on the boundary leads to a
certain, generally quadratic, $z$--dependent variation of the bulk
fields across the orbifold. This corresponds to coherent excitations of
the Kaluzu--Klein modes in the
orbifold direction. Let us consider an explicit example. We would like
to determine the $z$--dependence of the bulk fields that arises from the
explicit boundary potentials in~\eqref{Ui} proportional to
$v$. Inserting the corresponding part of the boundary stress energy
into eq.~\eqref{gt} and \eqref{ft} we find
\bea
 ds_5^2 &=& (\bar{g}_{\a\b}+\tilde{g}_{\a\b})dx^\a dx^\b
        = \left( 1+\frac{1}{3}\tilde{\f}\right)e^{-\bar{\b}}
          g_{4\m\n}dx^\m dx^\n +\left( 1+\frac{4}{3}
          \tilde{\f}\right)e^{2\bar{\b}}dy^2 \label{ftdw}\\
 \f &=& \bar{\f}+\tilde{\f} \\
 \tilde{\f} &=& -2e^{\bar{\b}-\bar{\f}}\e_{\rm DW}\left(
  z-\frac{1}{2}\right)\label{dw2}
\eea
where
\begin{equation}
 \e_{\rm DW} =-\frac{vR}{\sqrt{2}}\; .\label{edw}
\end{equation}
Note that the explicit potentials have the same height but opposite
sign on the two boundaries. Therefore, the $z$--dependent part of the
solution is proportional to the difference $(P_1-P_2)(z)$ of the
polynomials~\eqref{pol} and, hence, is
linear in $z$. This is to be contrasted to the general
case of unrelated boundary potentials which leads to a quadratic
variation. For constant moduli $g_{4\m\n}$, $\bar{\b}$ and $\bar{\f}$,
eq.~\eqref{ftdw}--\eqref{edw} represent the linearized
version of an exact BPS domain wall (three--brane) solution of
heterotic M--theory~\footnote{The exact solution of ref.~\cite{losw}
has the form $ds_5^2=a_0^2H\eta_{\m\n}dx^\m dx^\n +b_0^2H^4dy^2$,
$e^\f = b_0H^3$ where $H=c_0-\frac{2}{3}\e_{\rm DW}(z-\frac{1}{2})$
and $a_0$, $b_0$ and $c_0$ are constants. In fact, it also constitutes
an exact solution of the action~\eqref{S5}. Upon linearizing in
$\e_{\rm DW}$, setting $g_{4\m\n}\sim\eta_{\m\n}$ in \eqref{ftdw}
and appropriately matching the moduli this coincides with
\eqref{ftdw}--\eqref{dw2}.} in five dimensions that was found in
ref.~\cite{losw}. More precisely, the solution represents a pair of
domain walls stretched across the orbifold planes and it can be viewed
as the ``vacuum'' solution of the theory. At the same time, it is the
five--dimensional version of Witten's 11--dimensional linearized
background~\cite{w}. In this way, the four--dimensional orbifold
planes are identified with three--brane worldvolumes which carry the
observable (and hidden) low energy fields as their zero modes. In this
sense, our picture offers a concrete realization of the general idea
that the world arises from the worldvolume of a brane. Observe that
the size of the correction~\eqref{ftdw} is set by the dimensionless
quantity $\e_{\rm DW}$ in eq.~\eqref{edw}, which is just the product of
the boundary potential (measured in units of $\k_5^{-2}=M_5^3$) times
the size of the orbifold dimension. Therefore, the linearized
approximation that led us to the solution~\eqref{gt}, \eqref{ft} is
only sensible as long as $|\e_{\rm DW}|\ll 1$. We will discuss this in
more detail in the next section. 

By inserting the solution~\eqref{gsplit}--\eqref{ft} into the
action~\eqref{S5} and promoting the moduli $g_{4\m\n}$, $\bar{\b}$ and
$\bar{\f}$ to four--dimensional fields, we can now compute the first
order corrected action of the zero modes. We introduce the fields
\begin{equation}
 S=e^{\bar{\f}}\; ,\qquad T=e^{\bar{\b}}
\end{equation}
which are just the real parts of the ordinary $S$-- and
$T$--moduli~\footnote{The imaginary parts are absent because we have
omitted the corresponding fields in the five--dimensional
action~\eqref{S5} for simplicity. The complete reduction can be found
in ref.~\cite{losw1}.}. Note that $S$ measures the volume of the
internal Calabi--Yau space in units of $v_{\rm CY}$ whereas $T$
measures the size of the orbifold in units of $R$. We also introduce
the four--dimensional Newton constant
\begin{equation}
 G_N=\frac{\k_5^2}{16\p R}=\frac{1}{16\p RM_5^3}\label{GN}
\end{equation}
and the rescaled boundary fields
\begin{equation}
 C_i = \sqrt{\frac{M_5^3}{3}}\f_i\; .
\end{equation}
Then the four--dimensional effective action is given by
\bea
 S_4 &=& -\frac{1}{16\p G_N}\int_{M_4}\sqrt{-g_4}\left[
         R_4+\frac{3}{2}T^{-2}\pt_\m T\pt^\m T+\frac{1}{2}S^{-2}\pt_\m S
         \pt^\m S\right]\nn \\
      && -\int_{M_4}\sqrt{-g_4}\left[\frac{1}{2}\sum_{i=1}^2
         K_i\pt_\m C_i\pt^\m C_i+V_4\right] \label{S4}
\eea
with the ``K\"ahler metrics''
\begin{equation}
 K_i = \frac{3}{2T}\pm\frac{\e_{\rm DW}}{2S}
\end{equation}
and the four--dimensional potential
\begin{equation}
 V_4 = \frac{1}{2T^2}\sum_{i=1}^2V_{4i}+\frac{\e_{\rm
       DW}}{2ST}\sum_{i=1}^2\left(\pm\frac{2}{3}V_{4i}
       \pm\pt_{\bar{\f}}V_{4i}\right) +\frac{1}{T}RM_5^3V\label{V4}
\end{equation}
where the boundary potentials $V_{4i}$ normalized to four mass
dimensions are defined by
\begin{equation}
 V_{4i}=M_5^3V_i\; .
\end{equation} 
Note that the K\"ahler metrics and the potential have corrections
linear in $\e_{\rm DW}$ that originate from the domain wall.
Had we performed a simple truncation of the five--dimensional theory
by taking the bulk fields independent of the orbifold we would have missed
these corrections. What happened to the explicit potentials
proportional to $v$ in the five--dimensional action, eq.~\eqref{U} and
\eqref{Ui}? These potentials were actually responsible for the
existence of the domain wall solution in the first place. Performing a
reduction on this solution, they are canceled so that there is no
remnant potential for the $S$ and $T$ moduli in four dimensions.
Given the form of our five--dimensional action~\eqref{S5}--\eqref{Ui},
the bulk potential $V$ only depends on $S$ but not on $T$. The
four--dimensional effective potential~\eqref{V4} shows that, in this
case, the modulus $T$ cannot be stabilized. For the solutions based on
the four--dimensional effective action we, therefore, have to slightly
generalize our setup and assume that $V$ is a function of the orbifold
modulus $T$ as well. 

An alternative way to deal with the orbifold dependence of bulk fields
in a reduction to four dimensions is to keep all the Kaluza--Klein modes
instead of integrating them out. Recently, this has systematically
been carried out in ref.~\cite{ly}. In such a four--dimensional
action, the zero mode part does not have the above
corrections. Instead, there exist terms in the action which couple
Kaluza--Klein modes linearly to zero mode fields. Hence the
Kaluza--Klein modes cannot be set to zero consistently. In fact, solving
for these Kaluza--Klein modes would lead to the Fourier decomposition
of our orbifold--dependent corrections~\eqref{gt}, \eqref{ft}.

\vspace{0.4cm}

Besides being useful to calculate the four--dimensional action, the
method described above can also be used to find approximate solutions
of the five--dimensional theory. This is actually the main reason why
we have presented it here in some detail. Concretely, suppose one has
found a solution of the four--dimensional theory with
action~\eqref{S4}. Then, by
inserting this solution into eq.~\eqref{gsplit}--\eqref{ft}, we can
simply ``lift it up'' to obtain a solution of the five--dimensional
theory. Of course, for this solution to be sensible, the linearized
approximation that led us to eq.~\eqref{gt}, \eqref{ft} should be
valid. The condition for that to be true will be discussed in the
next section in some detail.

\section{Types of inflation and scales}

Which type of inflation for the action~\eqref{S5} do we want to
consider? First of all, in this paper we are interested in
``conventional'' potential--driven inflation rather than in a
pre--big--bang--type scenario~\cite{pbb}. Some solutions of
five--dimensional heterotic M--theory that might provide a basis for a
pre--big--bang scenario have been found in
ref.~\cite{cosm,cosm1,real}. Given that we want to focus on potential
energy, there are two obvious options.
\begin{itemize}

\item bulk potential energy: The potential energy is provided by the
bulk potential $V$ and the field $\f$ is the inflaton. This can also be
called ``modular inflation''.

\item boundary potential energy: The potential energy is due to the
boundary potentials $V_i$ (one of them or both) and the fields $\f_i$ are
the inflatons. This can be called ``matter field inflation''.

\end{itemize}

Clearly, in general, one could also have a mixture of both types. In
this paper, we concentrate on the second option of boundary potential
energy, which appears to be more interesting for a number of reasons.
Most importantly, the presence of the boundaries is the truely new
ingredient in the action~\eqref{S5}. Inflation from the boundary also
seems to be in accord with the current mainstream in four--dimensional
inflationary model--building~\cite{lr}. On the other hand, modular
inflation faces a number of problems associated with the steepness of
typical non--perturbative moduli potentials~\cite{bs,mod_infl}. From the
viewpoint of the recently proposed models~\cite{dim1} with a very
low ``fundamental'' scale $M_5\sim$ TeV, bulk inflation might not be
desirable because reheating from a bulk field might leave the
additional dimension too inhomogeneous to be consistent with
standard cosmology~\cite{add}. None of these arguments, of course,
disproves bulk inflation and it might still be an interesting
option. This will be discussed elsewhere.

\vspace{0.4cm}

Let us, from now on, concentrate on the case of potential energy on
the boundary. Our assumption is, hence, that the bulk potential
$V(\f )$ is zero or negligible for all field values $\f$ relevant
during inflation (which does not mean, however, that $V$ is identical to
zero). The potential energy is then dominated by the boundary
potentials $V_i$ and the four--dimensional effective potential is
given by
\begin{equation}
 V_4\simeq \frac{1}{2T^2}\sum_{i=1}^2V_{4i}+O(\e_{\rm DW})\; ,
           \qquad V_{4i}=M_5^3V_i\; .\label{V4s}
\end{equation}
In the previous section, we have given a method to obtain approximate
five--dimensional solutions from solutions of the four--dimensional
effective action. Therefore, if we can find a four--dimensional
inflationary solution (which we can if $V_4$ satisfies the usual slow
roll conditions), we can lift it up to a solution of the $D=5$
action. When is this effective four--dimensional approach sensible? In
the domain wall example of the previous section we have seen that the
condition
\begin{equation}
 |\e_{\rm DW}|=\frac{|v|R}{\sqrt{2}}\ll 1 \label{edwll}
\end{equation}
which involves the height $v$ of the domain wall potentials should
be satisfied in order for the linearized solution to be valid. 
In the context of heterotic M--theory, the relation~\eqref{edwll} is
required anyway since the formulation of the theory is not well known
beyond the linear level in $\e_{\rm DW}$. In any case, \eqref{edwll}
should be satisfied for the effective four--dimensional
action~\eqref{S4} to be sensible.

Inspection of the general linearized solution~\eqref{gt}, \eqref{ft}
shows that we should have a similar condition for the boundary
potentials $V_i$; namely, for
\begin{equation}
 \e_i\equiv V_iR=\frac{V_{4i}R}{M_5^3}
\end{equation}
we should have
\begin{equation}
 |\e_i|\ll 1. \label{linear}
\end{equation}
What is the interpretation of these conditions? We have generally
seen that boundary potentials (and any other form of boundary energy)
lead to inhomogeneities in the additional dimension or, in other
words, excite Kaluza--Klein modes associated with this dimension. If
$|\e_i|\ll 1$ this can be described in a linearized approximation,
either by integrating out the Kaluza--Klein modes at the linear level
as we have done to arrive at our four--dimensional action for the
zero modes, or by keeping the linearized Kaluza--Klein modes in
the four--dimensional action as in ref.~\cite{ly}. If, on the other
hand, $|\e_i|\gg 1$ this linearized approximation breaks down. Then the
simplest possibility is probably to work with the full
five--dimensional action. Hence, we distinguish two cases.
\begin{itemize}

\item $|\e_i|\ll 1$, the linear regime: We can solve the
four--dimensional effective action~\eqref{S4} for the zero modes and
lift up the solutions to five dimensions using the results of the
previous section. We will discuss this case in the following section.

\item $|\e_i|\gg 1$, the non--linear regime: We should solve the full
five--dimensional theory~\eqref{S5}. This case will be considered in
section 5 and 6.

\end{itemize}

What seems surprising about this criterion, at first, is that the
quantities $\e_i$ are linear in the size $R$ of the additional
dimension. The larger the additional dimension, the earlier the system
enters the non--linear regime. This can be understood as
follows. Consider the one--dimensional space associated with the
additional dimension. The orbifold planes appear as points sources in this
space. On the linear level, the fields generated by these point sources are
roughly described by the one--dimensional Green's function which is
simply the linear function $|y|$. The proportionality of $\e_i$ to the
orbifold size $R$ simply reflects this linear increase in $y$ of the
one--dimensional Green's function. This picture also suggests how the
above criterion should be modified for more than one additional
dimension. For example, for two additional dimensions the Green's
function is a logarithm and, hence, roughly a constant. Therefore,
$\e_i$ should be independent of the size of the additional
dimensions. Generally, for $d$ additional dimensions, one expects the
corresponding parameter $\e_d$ (suppressing the index $i$) to be given by
\begin{equation}
 \e_d =\frac{V_4}{M_{4+d}^{2+d}R^{d-2}}
\end{equation}
where $M_{4+d}$ is the fundamental scale of the $4+d$--dimensional
theory. Since one would generally require that $V_4\ll M_{4+d}^4$ and
$R^{-1}\ll M_{4+d}$ to have a field--theoretical description, it
follows that $|\e_d|\ll 1$ as long as $d>1$. We see, therefore,
that the case of one additional dimension, which is the relevant one for
heterotic M--theory, is special in that it is the only case where
inflation might take place in the non--linear regime.

\vspace{0.4cm}

We would like to be somewhat more explicit about scales in
order to get a feeling for when the above linearity criterion might be
satisfied. In the Ho\v rava--Witten context, one reference point is
the ``physical point'' at which the gauge and gravitational couplings
are matched~\cite{w}. At this point, one finds to lowest order
\bea
 \k^{-2/9} &\simeq& x\, 4\cdot 10^{16}\mbox{ GeV} \nn \\
 v_{\rm CY}^{-1/6} &\simeq& x\, 2\cdot 10^{16}\mbox{ GeV} \label{scales}\\
 R^{-1} &\simeq& x^3\, 2\cdot 10^{15}\mbox{ GeV}\nn
\eea
for the energy scales associated with the 11--dimensional theory, the
Calabi--Yau space and the orbifold. Here $x$ is a quantity of order
one which depends on the shape of the Calabi--Yau space and
parameterizes our ignorance of the precise relation between
the Calabi--Yau volume and the unification scale. 
Using eqs.~\eqref{M5} and \eqref{k5}, we find for the mass scale of the
five--dimensional theory
\begin{equation}
 M_5\simeq x\, 2\cdot 10^{17}\mbox{ GeV}\; .
\end{equation}
The linearity criterion~\eqref{linear} then translates into
\begin{equation}
 V_{4i}^{1/4}\ll x^{3/2}\, 6\cdot 10^{16}\mbox{ GeV}\; .\label{lin_spec}
\end{equation}
Hence, if during inflation the Calabi--Yau volume and the orbifold
radius are at the values that lead to coupling constant unification,
the Kaluza--Klein modes behave linearly as long as the boundary
potentials satisfy the bound~\eqref{lin_spec}. This has to be compared
with the COBE normalization which implies for the (four--dimensional)
inflationary potential that
\begin{equation}
 V_{4}^{1/4}\simeq\varepsilon^{1/4}\, 6.7\cdot 10^{16}\mbox{ GeV}\label{COBE}
\end{equation}
where
\begin{equation}
 \varepsilon = \frac{(V_4')^2}{2G_NV_4^2}
\end{equation}
is the usual slow roll parameter. The prime denotes the derivative
with respect to the inflaton. Comparison of eq.~\eqref{COBE} and
eq.~\eqref{lin_spec} shows an interesting coincidence of scales
suggesting that inflation might start just when the theory leaves the
non--linear regime. The bound~\eqref{lin_spec} is, however, very close
to the physical orbifold and Calabi--Yau scales
in~\eqref{scales}. Beyond energies of $v_{\rm CY}^{-1/6}$ the
effective five--dimensional theory breaks down. Also, beyond energies of
$\k^{-2/9}$ the description via 11--dimensional supergravity is no
longer viable. Clearly, beyond those scales our analysis does not
apply and there is no sense to talk about a non--linear regime as
defined above. Most likely, therefore, the theory undergoes a
transition from an M--theory regime directly into the linear regime.
Of course, the ``coincidence'' of scales is due to the fact
that the fundamental scales of our theory are much closer to the scale
suggested by COBE than, for example, the Planck scale. To
make this observation really meaningful one has to more closely analyze the
transition to the linear regime, something that
is beyond the scope of this paper. In any case, all these
statements relate to the ``physical point'' associated with the present
values of the Calabi--Yau volume and the orbifold radius. Clearly,
those values could have been different in the early universe, so that
we really do not know in which regime inflation took
place. Consequently, in the context of Ho\v rava--Witten theory, as
well as in a wider context, we should investigate both the
linear and the non--linear possibility.

\vspace{0.4cm}

Finally, we would like to discuss which part of the space should be
allowed to inflate from a ``phenomenological'' point of view. There are
two options. First, while the usual three--dimensional space
inflates the additional dimension is basically fixed. Secondly, both
the three--dimensional space and the additional dimension
inflate. While there is obviously no problem with the first option,
inflating the full space might lead to an additional dimension that is
too large. In particular, inflating it for the full $55$ or so e-folds
leaves one with a radius $R$ that would typically be some $24$ orders
of magnitude larger than the fundamental length scale $M_5^{-1}$ of
the theory. This is clearly unacceptable. It might be acceptable to inflate the
additional dimension for a short period (depending on the precise
value of the scales) and then stabilize it to its low energy
value. This stabilization of an inflating modulus might, however, be
hard to achieve theoretically. In this paper, we therefore favor the
first option of a non--inflating additional dimension.

\section{Linear case: The inflating domain wall}

In this section, we would like to construct inflationary solutions in the
linear case
\begin{equation}
 |\e_i|\ll 1\; .
\end{equation}
Following the general method presented in section 2, we should first
find an inflating solution of the four--dimensional effective
theory~\eqref{S4}. The existence of such a solution depends, of
course, crucially on the properties of this theory and its potential
in particular. Rather than going into model building, we will simply
assume suitable properties. Since we are interested in boundary
inflation, we assume that the moduli $S$ and $T$ have been
stabilized by the four--dimensional effective bulk potential~\eqref{V4}.
As discussed earlier, this requires a non--perturbative bulk potential
$V$ that depends on $S$ and $T$. To simplify formulae, we choose the coordinate
Calabi--Yau volume $v_{\rm CY}$ and the coordinate orbifold radius $R$
such that this stable point is at
\begin{equation}
 \bar{\f}=\ln S = 0\; ,\qquad \bar{\b}=\ln T = 0\; .
\end{equation}
Furthermore, since we would like the vacuum energy to be dominated by
the boundary potentials, we assume that the bulk potential vanishes
at this point; that is
\begin{equation}
 V(\bar{\f}=0,\bar{\b}=0)=0\; .
\end{equation}
Finally, we need our candidate inflatons, the boundary fields $C_i$, to
be slowly rolling. This requires the inequalities
\begin{equation}
 \frac{(\pt_{C_i}V_4)^2}{G_NV_4^2}\ll 1\; ,\qquad
 \frac{\pt_{C_i}\pt_{C_j}V_4}{G_NV_4}\ll 1
\end{equation}
to be satisfied for all $i,j=1,2$. Then, starting with the usual
four--dimensional metric
\begin{equation}
 ds_4^2=g_{4\m\n}dx^\m dx^\n =-d\t^2+e^{2\a_4}d{\bf x}^2
\end{equation}
for a spatially flat Robertson--Walker universe with scale factor
$\a_4$, the four--dimensional equations of motion from the
action~\eqref{S4} reduce to
\bea
 H^2&\equiv&\dot{\a}_4^2=\frac{8\p G_N}{3}V_4 \label{h2}\\ 
 \dot{C}_i &=& -\frac{1}{V_4}\pt_{C_i}V_4\; .\label{cd}
\eea
The usual inflating solution is then
\begin{equation}
 \a_4=H(\t -\t_0)
\end{equation}
with an arbitrary integration constant $\t_0$. The slow roll motion
$C_i(\t )$ of the inflatons can be obtained from eq.~\eqref{cd} once
an explicit potential has been specified. Note that with the
conventional relation~\eqref{h2} between the potential and the Hubble
parameter and using eq.~\eqref{GN}, the quantities $\e_i$ can be written as
$\e_i\sim H^2R^2$. The condition $\e_i\ll 1$ then corresponds to the
``intuitive'' criterion that the Hubble parameter $H$ should be
smaller than the mass $R^{-1}$ of the first Kaluza--Klein excitation.
In ref.~\cite{ly1} this criterion has been used to constrain the
Hubble parameter during inflation for TeV--scale gravity models.

We can now lift this solution up to five dimensions using the
formulae of section 2, in particular eq.~\eqref{gsplit}--\eqref{ft}.
We find~\footnote{A subtle point has to be taken into account if one
wants to explicitly verify that this solution satisfies the
five--dimensional equations of motion~\eqref{eomf}--\eqref{eomm} to
linear order. For the underlying four--dimensional solution, we have explicitly
assumed that the orbifold modulus $T=e^{\bar{\b}}$ has been
stabilized at $T=1$. The effective four--dimensional potential~\eqref{V4} shows
that this requires a $T$ dependence of the bulk potential
$V$. Since $\pt_TV_4(T=1)=0$ and $V(T=1)=0$,
one concludes from eq.~\eqref{V4} that
$\pt_{\bar{\b}}V(T=1)=R^{-1}\sum_{i=1}^2V_i+O(\e_{\rm DW})$. Moreover, in the
five--dimensional equations of motion we have not considered a $\b$
dependence of the bulk potential $V$. To incorporate such a case, the
potential $U$ in eq.~\eqref{eom55} has to be replaced by $U+\pt_\b V$.
Using this modification and the above expression for $\pt_\b V$ one
can indeed verify that the five--dimensional equations of motions
are satisfied.}
\bea
 ds_5^2 &=& -e^{2\n}d\t^2+e^{2\a}d{\bf x}^2+e^{2\b}dy^2\label{is}\\
 \a &=& H(\t -\t_0)+\tilde{\a} \\
 \n &=&\tilde{\a} \\
 \b &=& 4\tilde{\a} \\
 \f &=& \tilde{\f} 
\eea
where the orbifold dependent corrections $\tilde{\a}$ and $\tilde{\f}$
are given by
\bea
 \tilde{\a} &=& -\frac{\e_{\rm DW}}{3}\left( z-\frac{1}{2}\right)
                -\frac{R}{12M_5^3}\sum_{i=1}^2P_i(z)V_{4i}
                (\bar{\f}=0,C_i(\t )) \label{ait} \\
 \tilde{\f} &=& -2\e_{\rm DW}\left( z-\frac{1}{2}\right)
                +\frac{R}{2M_5^3}\sum_{i=1}^2P_i(z)\pt_\f
                V_{4i}(\bar{\f}=0,C_i(\t ))\; .\label{fit}
\eea
Recall that $P_i(z)$ are quadratic polynomials in the normalized
orbifold coordinate $z=y/R\in [0,1]$ that have been defined in
eq.~\eqref{pol}. In the metric~\eqref{is} we should, of course, only
consider terms linear in $\tilde{\a}$ and $\tilde{\f}$, in accordance
with our approximation.

Let us discuss the form of this solution. First we note that,
neglecting the very mild time dependence introduced by the slow roll
of the inflatons for the moment, the solution separates into a
time--dependent and an orbifold--dependent part. The time--dependent
part just corresponds to the inflationary expansion of the
three--dimensional universe. This expansion does
not occur just on one boundary, as one might naively expect,
but uniformly across the whole orbifold. To discuss the corrections,
let us concentrate on the scale factor $\a$. Although it expands at
the same Hubble rate $H$ everywhere across the orbifold, its actual
value depends on the orbifold point as specified in
eq.~\eqref{ait}. The first term in this equation is the familiar
linear contribution from the domain wall proportional to $\e_{\rm
DW}$. The second term arises from the boundary potentials and is
proportional to $\e_i$, as expected. It has a mild time--dependence
through the slow--roll change of the potentials. 

The bottom line of this section is, that the problem of finding
inflationary backgrounds in the linear regime can be adequately
approached in the four--dimensional effective action obtained by
integrating out the Kaluza--Klein modes. For our simple model, this
action is given in~\eqref{S4}. More realistic four--dimensional
effective actions from Ho\v rava--Witten theory can be found
in~\cite{hp,low,eff5,nse,susy5}. The full five--dimensional solution
is then obtained by lifting the four--dimensional
solution up, using the correspondence established in section 2.
This leads to the corrections~\eqref{ait} and \eqref{fit} corresponding to
Kaluza--Klein modes that are coherently excited by the non--vanishing
sources on the orbifold planes. One might also worry about other
excitations of the Kaluza--Klein modes unrelated to the orbifold
sources, such as remnants from an initial state. This
could be described by adding the tower of Kaluza--Klein modes to the
four--dimensional effective action~\eqref{S4}. Since we have
integrated out the orbifold sources, those modes would be free source--less
particles with masses $n/R$, where $n$ is an integer. During
inflation, oscillations
of these modes are simply damped away by the expansion. The condition
for this to happen efficiently coincides with our linearity criterion
$|\e_i|\ll 1$ and is, hence, satisfied. In the linear regime, the only
relevant excitations of Kaluza--Klein modes after a short period of
inflation are, therefore, the ones caused by the orbifold sources
computed above. As we will see, this changes in the non--linear regime
where $|\e_i|\gg 1$.

We would briefly like to mention some generalizations. It is clear
that the above method can be applied to other types of
four--dimensional cosmological solutions, for example to a preheating
solution with the energy density dominated by coherent
oscillations or to a radiation dominated solution, straightforwardly.
Basically, all one has to do is to replace the potentials in
eq.~\eqref{ait} by the appropriate energy density. What about the case
of more than one large dimension? In such a case, the variation across
the orbifold would not increase as a polynomial any more, as we have
seen earlier. Instead, for two additional dimensions we expect a
logarithmic behaviour and for more than two dimensions a power law fall--off.
Common to all these cases is, however, that energy density on the
four--dimensional plane coherently excites the bulk modes.

\section{Non--linear case: A solution by separation of variables}

We would now like to study the non--linear case; that is, we assume
\begin{equation}
 |\e_i|\gg 1\; .\label{non_linear}
\end{equation}
In this case, we should solve the full five--dimensional equations of
motion given in eq.~\eqref{eomf}--\eqref{eoml}. We will use the
boundary picture to do this. In general, we have two types of
potentials on the boundaries, namely the domain wall potentials and the
potentials $V_i$, corresponding to the two terms in \eqref{Ui}. We have
already stated that $\e_{\rm DW}$, the dimensionless quantity that
measures the strength of the domain wall corrections, should be
smaller than one in order for the M--theory description via
supergravity to be valid. The condition~\eqref{non_linear}, therefore,
states that the potentials $V_i$ will be dominating over
the domain wall. To simplify our problem, we will therefore
neglect the domain wall potentials. Certainly, there will be an intermediate
region between the non--linear and linear regime where both potentials are
significant. It will, however, be very difficult to find explicit
solutions in this regime. We, therefore, concentrate on the
case~\eqref{non_linear}. As a further simplification, let us assume
that the Calabi--Yau volume modulus $\f$ has been stabilized by the
bulk potential $V$; that is
\begin{equation}
 \f =\mbox{const}\; .
\end{equation}
Of course, we have to be careful that this assumption is consistent
with the boundary condition on $\f$, eq.~\eqref{eoml}. We have already
neglected the first term in this condition which originates from the
domain wall. The second term is related to the boundary potentials
$V_i$ and vanishes if those are taken to be independent of $\f$. We
will assume this in the following. In accordance with
our general assumption of boundary inflation, the potential energy
from the bulk potential should be negligible,
\begin{equation}
 V(\f )\simeq 0\; .
\end{equation}
Finally, we assume that the boundary potentials $V_i(\f_i)$ are
suitable slow--roll potentials so that the boundary fields $\f_i$ act
as the inflatons. Practically, this means that we treat $V_i$ simply as
constants. We recall from section 2 that the metric has the form
\begin{equation}
 ds_5^2 = -e^{2\n}d\t^2+e^{2\a}d{\bf x}^2+e^{2\b}dy^2
\end{equation}
where $\a$ and $\b$ are the scale factors of the three--dimensional
universe and the orbifold respectively. We also remark that, from
eq.~\eqref{alphab}, \eqref{betab}, the boundary conditions take the
form
\begin{equation}
 e^{-\b}\a '\left.\right|_{y=y_i}=e^{-\b}\n '\left.\right|_{y=y_i}
 =\mp\frac{1}{12}V_i\; .\label{bsimple}
\end{equation}
Even though the equations of motion~\eqref{eomf}--\eqref{eoml} have
now considerably simplified, they are still not easily soluble. The
reason is, of course, that we are dealing with partial
differential equations as opposed to ordinary ones that one usually
encounters in cosmology. The simplest strategy to solve partial
differential equations is separation of variables and this is what we
are going to do next. The general solution will be given in the
following section. 

To simplify the problem we first choose the conformal gauge $\n =\b$.
We will assume for the time being that the coordinate transformation
that led to conformal gauge leaves the boundaries at finite values of
the coordinate $y$. In this case, we can restore the conventions for our
coordinate system by shifting the boundaries back to $y=0,R$.
We are, then, looking for all solutions of \eqref{eomf}--\eqref{eoml}
(subject to the above assumptions) consistent with the separation ansatz
\bea
 \a &=& \a_0(\t )+\a_5(y) \\
 \b &=& \b_0(\t )+\b_5(y)\; .
\eea
The general solution to the equations of
motion~\eqref{eomf}--\eqref{eomr} in the boundary picture is
\bea
 \a &=& K(y\pm\t )+A \label{alpha}\\
 \b &=& K((1-\tilde{K})y\pm\tilde{K}\t )+B \label{beta}
\eea
where $K$, $\tilde{K}$, $A$ and $B$ are integration constants. We still have to
apply the boundary conditions~\eqref{bsimple}. This leads to
\begin{equation}
 K=\frac{1}{R}\ln\left(-\frac{V_1}{V_2}\right)\; ,\qquad
 \tilde{K} = 0\; ,\qquad
 B = \ln\left(-\frac{12K}{V_1}\right)\label{Ks}
\end{equation}
while $A$ remains arbitrary. For the arguments of the logarithms
to be positive, we have to further demand that $V_1$ and $V_2$ have opposite
signs such that $V_1+V_2<0$. It is not yet clear, whether these
restrictions on $V_i$ are general or whether they are related to our
choice of the coordinate system. We have assumed that in conformal
gauge the boundaries are at finite values. This need not
be the case if the coordinate transformation that led to conformal
gauge had a singularity. To cover such a case, we introduce a general
orbifold coordinate $\tilde{y}$ by
\begin{equation}
 y = \frac{1}{K}\ln (Kg(\tilde{y}))-B\; ,
\end{equation}
where $g(\tilde{y})$ is a monotonic, continuously differentiable
function for $\tilde{y}\in [y_1,y_2]$. Rewriting the
solution~\eqref{alpha}, \eqref{beta} in terms of this new coordinate
and applying the boundary conditions~\eqref{bsimple} leads to
\begin{equation}
ds_5^2 = -K^2g(\tilde{y})^2d\t^2+K^2g(\tilde{y})^2e^{\pm 2K(\t -\t_0)}
         d{\bf x}^2+g'(\tilde{y})^2d\tilde{y}^2\; ,\label{sg}
\end{equation}
where $g$ should satisfy at the boundaries
\begin{equation}
 g(y_1)=\mp\frac{12}{V_1}\; ,\qquad g(y_2) = \pm\frac{12}{V_2} \; .\label{gV}
\end{equation}
The upper (lower) sign applies to an increasing (decreasing) function $g$.
Here $K$ and $\t_0$ are arbitrary constants~\footnote{Choosing
$g(\tilde{y})=\frac{1}{K}-\tilde{y}$, we obtain a form similar to the
four--dimensional domain wall solution of~\cite{vil,is} and its
five--dimensional counterpart in~\cite{kl}.}.

\vspace{0.4cm}

\begin{figure}[h]
   \centerline{\psfig{figure=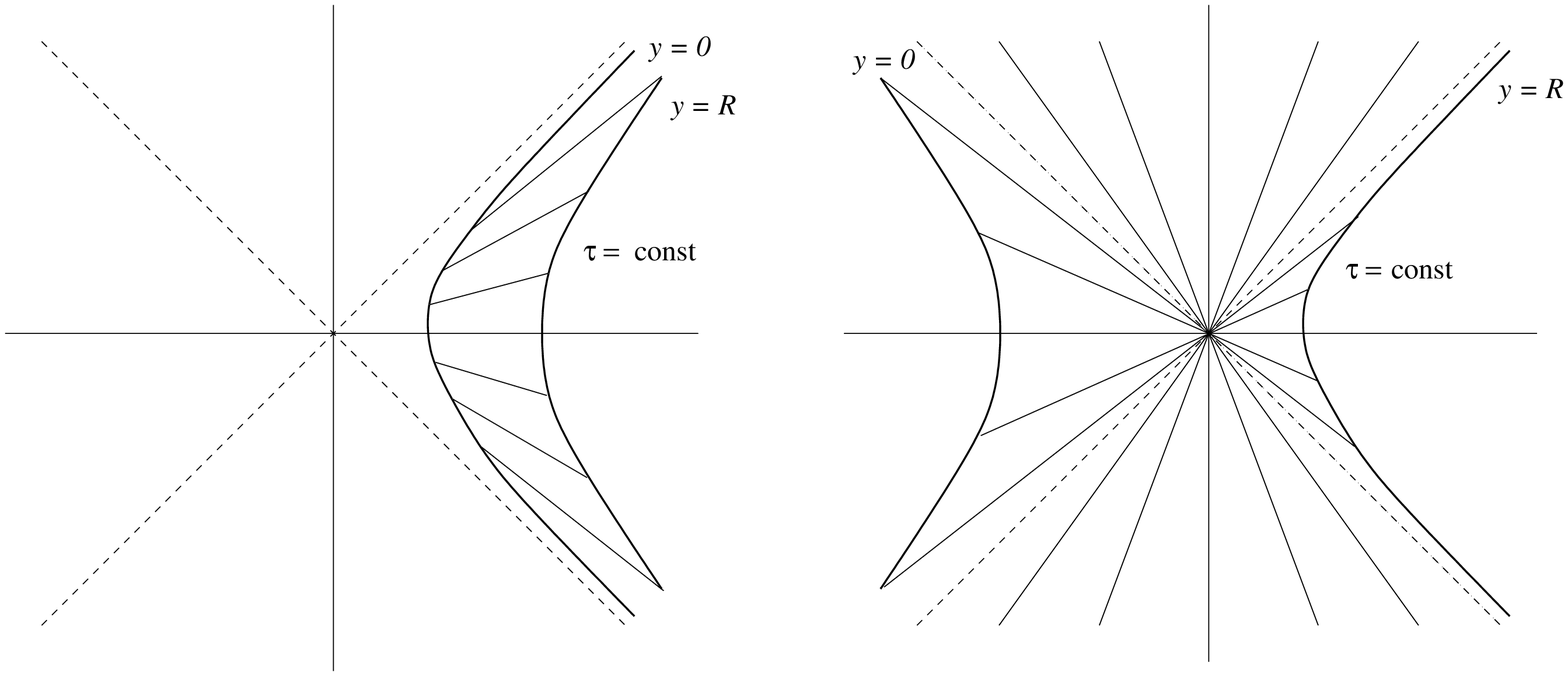,height=3in,width=6.8in}}
   \caption{Parts of Minkowski space corresponding to the solution in
            case 1 (left figure) and case 2 (right figure).}
   \label{fig1}
\end{figure}

Let us discuss some properties of this solution. As is well
known~\cite{vil,is}, the metric~\eqref{sg} is flat everywhere in the
bulk. What makes the metric nevertheless non--trivial is the presence of the
boundaries. While those boundaries are hypersurfaces with $y=$ const
in our coordinate frame, they would be mapped to de Sitter
hypersurfaces in coordinates where the metric~\eqref{sg} takes
Minkowski form. Indeed, if we define comoving times $t_i$ on each boundary by
setting $dt_i^2=K^2g(y_i)^2d\t^2$, the four--dimensional boundary
metrics take the form
\begin{equation}
 ds_{4,i}^2 = -dt_i^2+K^2g(y_i)^2e^{2H_i(t_i-t_{i0})}d{\bf x}^2
\end{equation}
with the Hubble parameters $H_i$ given by
\begin{equation}
 H_i = \pm\frac{V_i}{12}=\pm\frac{V_{4i}}{12M_5^3}\; .\label{Hi}
\end{equation}
The above linear relation between the Hubble parameters and the
potentials is quite unconventional. Usually the square of the Hubble
parameter is proportional to the potential, as in our linear case,
eq.~\eqref{h2}. We should distinguish two cases for the function $g$.
\begin{itemize}
\item $g(\tilde{y})\neq 0$ for all $\tilde{y}\in [0,R]$: In this case,
the metric~\eqref{sg} is regular everywhere on the orbifold, as in our
original form~\eqref{alpha}, \eqref{beta} of the solution in conformal
gauge. Since $g$ is continuous, $g(y_1)$ and $g(y_2)$ have to have
the same sign. From eq.~\eqref{gV}, it then follows that
\begin{equation}
 V_1V_2<0\; ,\qquad V_1+V_2<0\; .\label{c1}
\end{equation}
This are indeed the relations we found in conformal gauge above. 

\item $g(y_0)=0$ for some $y_0\in [0,R]$: In this case, the
metric~\eqref{sg} has a horizon at $y=y_0$. Now $g(y_1)$ has to
be negative and $g(y_2)$ positive (or vice versa) and we conclude
from \eqref{gV} that
\begin{equation}
 V_1>0\; ,\qquad V_2>0\; .\label{c2}
\end{equation}
This case could not be obtained from the solution in conformal gauge.
\end{itemize}
In fig.~\ref{fig1} two--dimensional Minkowski space
corresponding to the $(\t ,y)$ plane is depicted. We have indicated
the portions of this space that correspond to the above two types of solutions.
As explained earlier, if the metric~\eqref{sg} is
written in Minkowsi form by applying an appropriate coordinate
transformation the $y=$ const orbifold planes are mapped into de
Sitter hypersurfaces. In the two--dimensional pictures they appear as
hyperbolas. The space between those hyperbolas in fig.~\ref{fig1}
represents the orbifold and the lines indicate the locations of constant 
time, $\t =$ const. In case 1 (left figure \ref{fig1}), both bondaries are
on the same side of the light cone. Signals that travel in the bulk
will always reach the boundary after a finite time. In particular, a
signal sent from one plane will always reach the other one in a
finite time. These causal properties are somewhat counterintuitive
in that one would expect the existence of signals that travel
exclusively in the bulk without ever hitting a boundary. The figure
shows that, in fact, such signals do not exist. On the other hand, in
case 2 (right figure \ref{fig1}), a signal emitted from one of the
boundaries will never reach the other one. In this sense the two
boundaries are causally decoupled. Again, this property is somewhat
unexpected intuitively.\\ 

The physical length of the orbifold (obtained by integrating
$\sqrt{g_{55}}$ over the coordinate interval $[0,R]$, as usual) is
static and in both cases given by
\begin{equation}
 R_{\rm phys}=12\left(\frac{1}{V_1}+\frac{1}{V_2}\right)\; . \label{Rphys}
\end{equation}
This is positive due to the conditions~\eqref{c1} and \eqref{c2}, as
it should. In view of eq.~\eqref{Hi}, this leads to
\begin{equation}
 H_i\sim \frac{1}{R_{\rm phys}}
\end{equation}
for potentials $V_i$ of the some order of magnitude.
This relation is quite promising, since it directly relates the Hubble
parameters to the size of the additional dimension. We would like to
point out that, unlike in the linear case, we did not assume the
non--perturbative bulk potential $V$ to depend on the orbifold
modulus. Nevertheless, the orbifold size $R_{\rm phys}$ turns out to be
time--independent. Moreover, it is fixed by eq.~\eqref{Rphys} in terms
of the boundary potentials.

We have already mentioned that the above solution is the
only separating solution compatible with our initial assumptions of a
stabilized modulus $\f$ and slow roll of the boundary fields
$\f_i$. There is yet another sense in which this solution is
unique. Suppose one is interested in solutions where the bulk moduli
fields $\b$ and $\f$  are constant in time or slowly moving (with a
negligible contribution from the bulk potential $V$ to the vacuum
energy), the boundary fields are slowly rolling in their potentials
$V_i$ and the Hubble parameter $H\equiv\dot{\alpha}$ changes only
slowly in time. Practically, one can then neglect terms containing
$\dot{\b}$, $\dot{\f}$, $\dot{\f_i}$ and $\dot{H}$ in the equations of
motion. Heuristically, these properties are what one expects
from an inflating solution in five dimensions, in analogy with the
four--dimensional case. Then, one can show that all solutions with
these properties are approximated (in the sense that slow--roll
corrections have been neglected) by~\eqref{alpha}--\eqref{Ks}.
To do this, one does not need the technical assumption of separability
that we have used so far. In this sense, we have found the unique
solution with boundary inflation in five dimensions.

Can this solution, then, be used as the basis for an inflationary model
in five dimensions? We have to keep in mind that we have not solved
the equations of motion in general yet, but rather found a specific
solution by imposing separability or, equivalently, ``reasonable''
physical conditions of what an inflationary solution should look like.
Therefore, our solution might be very special in the sense that it,
perhaps, can only be obtained from a set of initial conditions with
measure zero. In other words, the solution could be unstable against
small perturbations. We will analyze this question in the following section. 

\section{Non--linear case: General solution}

In this section, we present the general solution. We start with the
same setup as in the beginning of section 5. Since we are in the
non--linear regime
\begin{equation}
 |\e_i|\gg 1
\end{equation}
we can neglect the domain wall potentials. Furthermore, we assume that
$\f$ has been stabilized by the bulk potential $V$ at a point with vanishing
potential energy $V(\f )=0$ (also, as a technical assumption, we have
to require the boundary potentials to be independent on
$\f$). Furthermore, the boundary potentials $V_i$ should lead to slow
roll of the fields $\f_i$. We will, however, not assume separability
or slow time evolution of any other field.

\vspace{0.cm}

It is useful to introduce light--cone coordinates
\begin{equation}
 x^{\pm}=\t\pm y
\end{equation}
and rewrite the equations of motion~\eqref{eomf}--\eqref{eomr} in
terms of these coordinates choosing conformal gauge $\n =\b$. Using
the simplifications that follow from the above setup, one finds
\bea
 \pt_+^2\a -2\pt_+\a\pt_+\b +\pt_+\a^2 &=& 0\label{lc00}\\
 \pt_-^2\a -2\pt_-\a\pt_-\b +\pt_-\a^2 &=& 0\label{lc55}\\
 \pt_+\pt_-\a +3\pt_+\a\pt_-\a &=& 0 \\
 2\pt_+\pt_-\a +\pt_+\pt_-\b +3\pt_+\a\pt_-\a &=& 0\; .\label{lca}
\eea
The boundary conditions~\eqref{alphab} and \eqref{betab} specialize to
\begin{equation}
 e^{-\b}\a '\left.\right|_{y=y_i}=e^{-\b}\b '\left.\right|_{y=y_i}
 =\mp\frac{V_i}{12}\; ,\label{bcond}
\end{equation}
where the upper (lower) sign applies to the boundary at $y=y_1$
($y=y_2$). The equations of motion~\eqref{lc00}--\eqref{lca} are quite
similar (although not identical) to those of two--dimensional dilaton
gravity~\cite{callan} with vanishing cosmological constant. In fact,
\eqref{lc00}--\eqref{lca} can be obtained from the two--dimensional
action
\begin{equation}
 S_2 = -\int\sqrt{-g_2}e^{3\a}\left[ R_2-24\, \pt_a\a\pt^a\a\right]
\end{equation}
with the two--dimensional metric in conformal gauge given by
$g_{ab}=e^{2\b}\eta_{ab}$. Here we have used indices $a,b,\dots =0,5$
for the space $(\t ,y)$. Using the methods of ref.~\cite{callan}, we
can find the following general solution of~\eqref{lc00}--\eqref{lca} 
\begin{equation}
 \begin{array}{lllllll}
  \a &=&\frac{1}{3}\ln u&,\qquad&\b &=&w-\frac{1}{3}\ln u \\
  u &=& u_+(x^+)+u_-(x^-)&,\qquad&w &=&w_+(x^+)+w_-(x^-)
 \end{array}
\end{equation}
where $u$ and $w$ are free fields, as indicated. The ``left-- and
right--movers'' $u_\pm (x^\pm )$ and $w_\pm (x^\pm )$ are not
completely independent but, rather, subject to certain relations that
originate from the constraint equations~\eqref{lc00} and \eqref{lc55}.
Three cases can be distinguished
\begin{itemize}

\item Case 1: If $\pt_+u_+\neq 0$ and $\pt_-u_-\neq 0$ then
\begin{equation}
 w_\pm = \frac{1}{2}\ln (\pt_\pm u_\pm )+C_\pm\; .
\end{equation}

\item Case 2: If $\pt_+u_+\neq 0$ and $\pt_-u_- = 0$ then
\begin{equation}
 w_+ = \frac{1}{2}\ln (\pt_+ u_+ )+C_+\; .
\end{equation}
and $w_-$ is arbitrary.

\item Case 3: If $\pt_+u_+ = 0$ and $\pt_-u_- \neq 0$ then
\begin{equation}
 w_- = \frac{1}{2}\ln (\pt_- u_- )+C_-\; .
\end{equation}
and $w_+$ is arbitrary. 

\end{itemize}
Here $C_\pm$ are arbitrary integration constants. In each of the three
cases, we still have two arbitrary functions at our disposal. While
this constitutes the most general solution of the equations of
motion~\eqref{lc00}--\eqref{lca}, we have not yet taken into account
the boundary conditions~\eqref{bcond}. As we will see, this determines
one of the functions and imposes a periodicity constraint on the other
one. Before we come to that, we observe that in case 1 we have the
relation
\begin{equation}
 \b = 2\a +\frac{1}{2}\ln (9\pt_+\a\pt_-\a )+C_++C_-\label{cc1}
\end{equation}
between the scale factors $\a$ and $\b$. Suppose that we have found a
solution with an inflating three--dimensional universe in this
case. Neglecting $y$--dependence, the scale factor $\a$ is then
roughly given by $\a\sim H\t$ where $H$ is the Hubble parameter.
In this case, the second term on the right hand side of eq.~\eqref{cc1} is
approximately constant since $\pt_\pm\a$ should be related to the Hubble
parameter. Hence, up to an overall normalization we have $\b\sim
2\a$. This shows that, without assuming any initial fine--tuning, at
the end of inflation, the orbifold orbifold has expanded twice as much as the
three--dimensional universe. For the reasons discussed at the end
of section 3 we, therefore, disregard this possibility. We remark
that solving the boundary condition~\eqref{bcond} for case 1 leads
to periodicity constraints in terms of elliptic functions. 

Let us now turn to cases 2 and 3. Fortunately, the boundary conditions
can be explicitly solved in these cases. We find that $\a$ and $\b$
can be expressed in terms of a single real function $f$ as
\bea
 \a &=& \frac{1}{3}\ln f(x^\pm) \label{a}\\
 \b &=& \ln\left[\frac{4(f'(x^\pm ))^{1/2}(f'(x^\mp ))^{1/2}}
                 {(\mp V_1)(f(x^\mp ))^{2/3}(f(x^{\pm}))^{1/3}}\right]\label{b}
\eea
where $f'$ denotes the derivative of $f$ and, as usual, $x^\pm =\t\pm
y$. The function $f$ should have the periodicity property
\begin{equation}
 f(x+2R)=\left[e^{\mp 2RK}(f(x))^{-1/3}+\tilde{k}\right]^{-3}
 \label{period0}
\end{equation}
for all $x$, where
\begin{equation}
 K=\frac{1}{R}\ln\left( -\frac{V_1}{V_2}\right) \label{K1}
\end{equation}
and $\tilde{k}$ is a constant. The condition~\eqref{period0} can be
solved in terms of a periodic function $p(x)$ satisfying
\begin{equation}
 p(x+2R)=p(x)
\end{equation}
for all $x$. One finds
\bea
 f(x) &=& \left[ p(x)e^{\mp Kx}+k\right]^{-3}\qquad\mbox{for }K\neq 0 
          \label{opt1}\\
 f(x) &=& \left[ kx+p(x)\right]^{-3}\qquad\qquad\mbox{for }K=0\label{opt2}
\eea
where $k$ is another constant. The upper sign in the above solutions
always corresponds to case 2, the lower one to case 3. The definition
of $K$, eq.~\eqref{K1}, shows that the boundary potentials need to have
opposite sign for the solution to exist. We are, therefore, in the
first case~\eqref{c1} of the previous section which is as expected
since we have used conformal gauge. We will stick to this case, for
simplicity, in the following. The analog of the second case~\eqref{c2}
can again be obtained by employing a more general coordinate
system. Our main conclusions apply to this case as well.
Furthermore, the periodic
function $p(x)$ and the constant $k$ have to be chosen such that the
logarithms in \eqref{a} and \eqref{b} are well--defined. Apart from
those restrictions, $p(x)$ and $k$ are
arbitrary. Eq.~\eqref{a}--\eqref{opt2} is the most general solution of
the system~\eqref{lc00}--\eqref{bcond} for the cases 2 and 3 which, as
we have seen, are the interesting ones in the present context. Since
we can more or less freely choose one periodic function we have, in
fact, found a very large class of solutions.

Depending on the value of $K$ we have two different possible forms of
the function $f$, given in eq.~\eqref{opt1} and \eqref{opt2}. The
second option, $K=0$, is realized for a vanishing ``total potential
energy'', $V_1+V_2=0$. Not surprisingly, in this case, the scale
factor $\a$ does not inflate but (roughly) shows a power law
behaviour, as can be seen by inserting~\eqref{opt2} into
eq.~\eqref{a}. Consequently, this second option is only of limited
interest to us and we will concentrate on the first one, $K\neq 0$.
This case should contain our simple separating
solution~\eqref{alpha}--\eqref{Ks} of the previous section. Indeed,
if we choose
\begin{equation}
 p(x) = \mbox{const}\; ,\qquad k=0
\end{equation}
in eq.~\eqref{opt1} we recover this solution. What happens for other
choices? Let us consider the case $k\neq 0$ and $p(x)$ periodic but
otherwise arbitrary. If the argument of the
exponential in eq.~\eqref{opt1} is negative (the case that would lead
to an inflationary expansion if $k$ were zero) then, after a very
short time $f$ will be approximately constant, unless $k$ is
exponentially small. Inserting such an approximately constant $f$ into
the expressions~\eqref{a}, \eqref{b} for the scale factors $\a$ and
$\b$ shows that the three--dimensional universe becomes
static while the orbifold collapses. In the opposite case, where the
argument of the exponential is positive, $f$ falls exponentially and,
hence, from eq.~\eqref{a}, the three--dimensional universe collapses.
Either way, to have a stable configuration, we should choose $k=0$.
In this case, the scale factors read explicitly
\bea
 \a &=& \pm Kx^\pm -\ln p(x^\pm) \label{a1}\\
 \b &=& Ky+\ln\left[\left(\frac{\mp 12}{V_1}\right)
        \left(\pm K-\frac{p'}{p}(x^\pm )\right)^{1/2}
        \left(\pm K-\frac{p'}{p}(x^\mp )\right)^{1/2}\frac{(p(x^\mp ))^{1/2}}
        {(p(x^\pm ))^{1/2}}\right]\; .
\eea
We still have the freedom to choose the periodic function $p$. In the
above expressions this function always depends on $x^\pm =\t\pm
y$. It, therefore, leads to an oscillation in time with a period $2R$.
If we do not want the scale factor $\a$ to be significantly effected
by this, we should choose the maximum amplitude of $p$ to be sufficiently
small, so that $p(x)\simeq$ const. This basically brings us back to
our separating solution of the previous section. In essence, this
solution is the only one with the desired properties within our setup.

The discussion of this section has revealed two serious problems
with the separating solution. First of all, it corresponds to a very
specific choice of initial conditions satisfying either $k=0$ or $k$ being
exponentially small. All other values of $k$ lead to a collapsing
solution. This implies that the case $k=0$, as it stands, corresponds
to an unstable situation. A small perturbation that leads to a
non--vanishing $k$ will cause a collapse of the universe. 
The other problem is related to the presence of the periodic function
$p$. This function, in fact, encodes the information about the initial
inhomogeneity in the orbifold direction and this inhomogeneity
survives the whole period of inflation. Of course, this is related to
the fact that we are not inflating the orbifold as well which would
dilute those inhomogeneities. In the effective four--dimensional
linear case of section 4, oscillations of Kaluza--Klein modes were
damped away quickly due to the inflationary expansion. Apparently
this is no longer true in the non--linear five--dimensional regime.
Although those modes do not affect the homogeneity of the
three--dimensional universe directly, they could potentially be
harmful. For example, eq.~\eqref{a1} shows that the ``Hubble
parameter'' $\dot{\a}$ contains the periodic function $p(\t\pm y)$
and consequently oscillates in times. Hence, the modes could have
some influence on density fluctuations. Also, their
eventual decay into gravitons could leave unwanted relics. In any
case, the presence of the function $p$ contradicts somewhat the
philosophy of inflation which is supposed to wipe out any initial
information. 

Is there a possible cure for the stability problem? So far, we did not
attempt to stabilize the orbifold in any way. It is clear that this
fact is related to the instability that we encounter. While stabilizing the
orbifold modulus in a four--dimensional effective theory by simply
inventing a potential is relatively straightforward (although
understanding the origin of this potential is not), this is not the case
in five dimensions. The orbifold size in five dimensions is basically 
measured by the component $g_{55}$ of the metric. A bulk potential for
this component would break general five--dimensional covariance in a
very strong way. Also, since $g_{55}$ typically varies across the
orbifold it could not be everywhere in a minimum of that
potential. Alternatively, one could postulate a bulk potential that
only depends on the zero mode of $g_{55}$; that is, the length of the
orbifold. This would break general covariance not more seriously than
it already is broken by the presence of the orbifold. Suppose such a
potential had a minimum for a certain orbifold length $R_0$. Could an
orbifold modulus stabilized at this minimum be made consistent with
our solution? The apparent problem is that the orbifold size is
already fixed by eq.~\eqref{Rphys} in terms of the boundary potentials
$V_i$. There is no obvious reason why $R_0$ and $R_{\rm phys}$ should,
a priori, coincide. If we assume they do, for some reason, at a
certain time, this situation could only be maintained if the boundary
inflatons slow--roll in a very specific way so as to leave
\begin{equation}
 R_0=12\left(\frac{1}{V_1(\f_1)}+\frac{1}{V_2(\f_2)}\right)
\end{equation}
unchanged. This would correspond to a strong correlation of the motion
of the two inflatons. Without a detailed analysis
of the dynamics, which probably has to be carried out numerically, it
is hard to tell whether this would actually happen or whether, instead, the
orbifold modulus would start to strongly oscillate around its minimum
thereby destroying inflation.

Another way to stabilize the orbifold which avoids these problems is
to have a potential for $g_{55}$ on the boundary. Although such a
potential cannot appear
directly in the boundary actions in~\eqref{S5}, it may appear in one of
the Bianchi identities of heterotic M--theory in five
dimensions~\cite{hor,gau,eff5} which contain sources located on the
orbifold plane. Particularly, a potential from
gaugino condensation would manifest itself in the Bianchi identity.
Generically, however, it seems to be difficult to stabilize moduli
with potentials from gaugino condensation, in particular in the context of
cosmology~\cite{bs} (see however~\cite{beatrice}). In ref.~\cite{choi}
stabilization of the orbifold was achieved by a combination of gaugino
condensation and other non--perturbative effects resulting from
internally wrapped membranes. Although worth investigating in our
context, all those options go beyond our simple toy model and will not
be explicitly considered here.

To summarize, while there seem to be interesting
solutions with boundary inflation in the non--linear,
five--dimensional regime, a
closer investigation shows that they have problems related to the
stabilization of the orbifold and to inhomogeneities in the orbifold
direction that are not diluted. The stabilization problem is, of
course, very general and we should not be surprised to encounter it in
our cosmological context. It is conceivable that whatever eventually
stabilizes the orbifold, also saves our non--linear cosmological
solution. While the persistence of the orbifold inhomogeneities
appears to be in contradiction with the inflationary paradigm they
might, under certain conditions, be acceptable and give rise to
interesting predictions. It certainly needs more work to finally
decide these questions. Probably one also has to go beyond the simple
model we have used in this paper.

\vspace{0.4cm}

{\bf Acknowledgments}
A.~L.~would like to thank Bruce Bassett, Arshad Momen and Graham Ross
for helpful discussions. A.~L.~is supported by the European Community
under contract No.~FMRXCT 960090. B.~A.~O.~is supported in part by 
DOE under contract No.~DE-AC02-76-ER-03071 and by a Senior 
Alexander von Humboldt Award. D.~W.~is supported in part by
DOE under contract No.~DE-FG02-91ER40671. 
 


\end{document}